\documentclass[9pt,twocolumn,twoside]{osajnl}

\journal{ao} 

\setboolean{shortarticle}{false}

\title{An Arduino-based, low-cost imaging incubator for extended live cell imaging}

\author[1,*]{Vincent M. Rossi}
\author[1]{Katherine C. Davidson}
\author[2]{Lauren E. Moore}

\affil[1]{Department of Physics and Astronomy, Washburn University, 1700 SW College Ave, Topeka, KS 66621, USA}
\affil[2]{Department of Biology, Washburn University, 1700 SW College Ave, Topeka, KS 66621, USA}

\affil[*]{Corresponding author: vincent.rossi@washburn.edu}

\begin{abstract}
In order to image live cells for prolonged periods of time, an Arduino-based, low-cost imaging incubator was constructed. The imaging incubator keeps cells viable by controlling for temperature and $CO_2$ in order to maintain physiological conditions for cells during imaging. All devices and parts employed in the build were typical maker-type components in order to minimize the cost of the imaging incubator. The imaging incubator allows for real-time imaging of live cells exposed to any desired perturbation or stimulus. As a proof of the system's functionality, cells are imaged over 24 hours while remaining viable in the imaging incubator.
\end{abstract}

\setboolean{displaycopyright}{true}

\begin{document}

\maketitle


\section{Introduction}
Keeping living cells viable while imaging over moderate time intervals can be precarious, let alone for an extended period of time.\cite{natBiotech} The authors were interested in imaging the real-time response of mammalian cells exposed to different stimuli for up to 24 hours. Such prolonged time frames for imaging live cells should permit the monitoring of cells through the cell cycle while exposed to different stimuli. The desire to image mammalian cells for an entire day necessitates the need to keep cells viable over that time frame via regulation of temperature and $CO_2$. As a response to this need, the authors have created an Arduino-based, low-cost imaging incubator designed to keep cells viable for prolonged periods of time while held on the sample stage of a custom, bench-top imaging device.

Real-time, prolonged imaging of cells affords researchers dynamic information, including changes in cell morphology, responses to various stimuli or reagents, cell proliferation, motility and cell tracking.\cite{flLive,jove} Retail devices that are available for prolonged, live cell imaging are generally of two sorts: stages that can be added to any microscope or complete imaging systems. Stages that can be added to any microscope have the advantage of being utilized for any imaging application needed, across various platforms. Environmental control stages that can be used on any microscope start around \$450, but those less expensive units only control for temperature and not $CO_2$.\cite{amscope} Stages that can be added to any microscope that also control for $CO_2$ have starting prices above \$12,000.\cite{wpi}

Complete imaging systems that allow for environmental controls come in two varieties: those that can be placed within an existing incubator and those that have an incubation system incorporated onto their stages. Systems that are compact enough to insert into an incubator start around \$20,000 and can have prices above \$40,000.\cite{phi,juli} These systems also have the disadvantages of taking up the majority of incubator space and possibly exposing neighboring cells in the incubator to undesired light. To negate these disadvantages, these units would require their own, dedicated incubator, therefore adding thousands of dollars to their total cost. As an alternative, there are many bench-top microscopes available on the market with incubator stages integrated into their design. While these scopes don't take up incubator space, many are not priced as complete systems, requiring add-on units in order to achieve live cell imaging. In the worst case scenario, having to purchase additional units in order to achieve live cell imaging can double the units' base pricing. These scopes start at around \$12,000 and the complete systems can run over \$50,000.\cite{luma,evos}

The imaging incubator presented here was designed and constructed using a minimal research budget. In order to keep costs low, all devices, components and supplies used in the construction of the imaging incubator were acquired from typical maker sources, hardware stores and online retailers. This begins with the use of an Arduino Uno\cite{arduino} in place of a typical DAQ for interfacing with devices. All of the hardware, supplies, components and parts needed to build the imaging incubator were purchased for under \$350. At this price, the imaging incubator presented in this paper starts at less than 10\% of the cost of sample stages and closer to 1-2\% of the cost of complete imaging systems on the market with similar features, affording researchers a savings of thousands to tens of thousands of dollars.

While the Arduino IDE\cite{arduinoIDE} could be employed for monitoring and control of temperature and $CO_2$ levels, the authors instead relied upon LabVIEW (National Instruments, Austin, TX) for device control. This decision was made because of the need to also control optical and imaging devices associated with the custom microscope itself. By writing the imaging incubator control program via LabVIEW, both the imaging incubator and imaging device control are integrated into a single program. Software costs are not included in the cost of the imaging incubator as the lab already had a license for LabVIEW. In lieu of using LabVIEW, the open source Arduino IDE could be implemented for control of the imaging incubator at no cost.

The remainder of this paper covers the design and implementation of the Arduino-based imaging incubator. In Sec. \ref{system}, the design and build of the imaging incubator is detailed. This includes the enclosure itself, along with the various control systems and hardware. The software used for controlling the different components and systems is covered in Sec. \ref{software}. Finally, to demonstrate the imaging incubator's functionality, sample images of mammalian cells acquired via the imaging incubator over 24 hours are presented in Sec. \ref{results}, before giving concluding remarks (Sec. \ref{conclusion}). A list of all parts and components is summarized in the Appendix, along with block diagrams from the LabVIEW program used to automate the imaging incubator. The latter is also available via the article's online supplemental materials.

\section{System Design}
\label{system}

\begin{figure}[htb]
\centering\includegraphics[width=8.5cm]{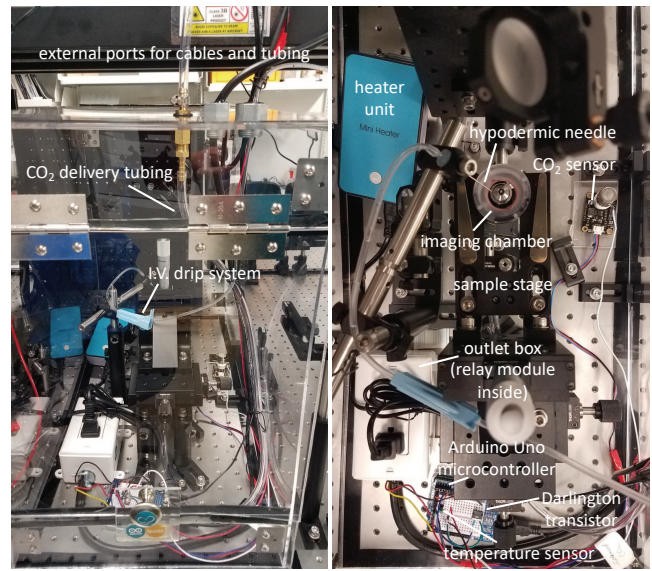}
\caption{The imaging incubator as viewed from the front (left pane), inside (middle pane) and top (right pane). Relevant components are labeled.}
\label{pic}
\end{figure}

While a compact, stage-top design would have been preferred, this led to issues providing heat from an external unit to a compact space. For ease of design, the imaging incubator was therefore designed as an enclosure over a portion of the custom microscope that would also completely enclose the heating unit. The imaging incubator (Fig. \ref{pic}) consists of the following elements: the enclosure, an Arduino Uno microcontroller for interfacing, temperature and $CO_2$ control systems and a fluid system consisting of a drip system for replenishing media for cells as it evaporates. These elements and systems will be covered in this section, while software for controlling the system devices will be discussed in Sec. \ref{software}. A manually controlled UV lamp (CTUV-6WTIMERUVC, Aopu Lighting Co) is also installed at the top of the enclosure for sterilization purposes. Sterile tissue culture practices were observed when using the imaging incubator, as would be done for any incubator.

\subsection{Enclosure}
The enclosure ($24'' \times 11 \: 1/2'' \times 18''$) for the Arduino-based imaging incubator was built around the sample stage of a custom built imaging system. In order to fully enclose the sample, additional optical elements needed to be enclosed within the imaging incubator as well. The enclosed optical elements included mirrors for redirection of the incident light and the microscope objective. The remainder of the microscope's optical elements and components were all outside of the imaging incubator, including the light source, collimating and imaging lenses and camera. Clear acrylic plexiglass sheets (1/8'' thick) were used for the four sides and lid of the imaging incubator. Opaque or black acrylic sheets could have been used in lieu of clear sheets, but since the microscope's camera was not housed within the imaging incubator, light contamination would have still been an issue for the imaging system. Whereas the use of an opaque enclosure would necessitate the need for cutting ports and then resealing them with glass windows for light input and output, a translucent enclosure eliminated the need to cut optical ports through the enclosure. The transparent enclosure also permits the imaging incubator to have uses for additional, future imaging devices which could have different incoming and outgoing beam paths.

Right-angle brackets were used in order to bolt the sides of the enclosure together. The front panel was cut at the top and bottom in order to create a door for access. The top of the front panel and door were joined together with cabinet hinges and a cabinet knob was added to the front of the door for access. The bottom portion of the front panel and the door were outfitted with a magnetic cabinet door catch in order to hold the incubator door closed when in use. Two through holes were drilled in the top sheet prior to assembly and pair of 1/2'' water tight conduit connectors were mounted in order to serve as tight fittings through which to pass cables and tubing. A third through hole was drilled for installation of a 1/4'' hose barb in order to port $CO_2$ via tubing. To improve the seal, o-rings were used at the junctions between the through holes in the cover and the conduit connectors and hose barb.

The enclosure was placed atop the optical bench, with strips of silicone rubber sheets (1/16'' thick) placed in between the bench top and enclosure in order to create an airtight seal. Additional strips of silicone rubber were placed inside the imaging incubator across the seams between the incubator door and enclosure in order to improve the seal at the seams, although this only provided a small improvement. Strips of rubber sheets were also placed atop the vertical walls of the enclosure to create a seal with the top panel, again made from the same acrylic sheet as the walls of the enclosure. A bead of caulking was also used to join the gaskets and enclosure walls and the corners of the enclosure where the walls met in order to ensure the best seal possible for the imaging incubator. Caulking was also used to seal the gaps between the conduit connectors, cables and tubing.

\subsection{Temperature \& $CO_2$ Control Systems}
In lieu of using a DAQ in order to interface between the controlling computer and devices associated with the imaging incubator, an Arduino Uno microcontroler was used as a low-cost alternative. The imaging incubator combines multiple systems that can be found via online Arduino and maker communities.\cite{outlet, CO2, solenoid}

\begin{figure*}[htb]
\centering\includegraphics[width=17cm]{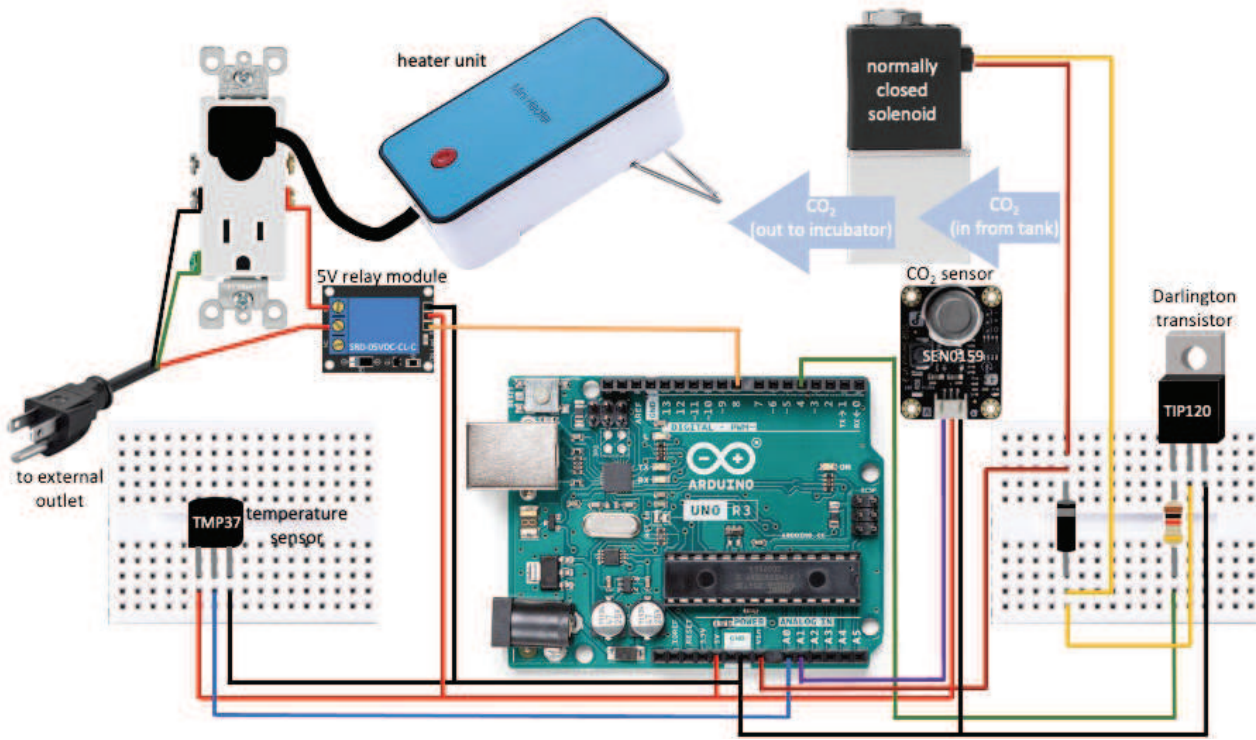}
\caption{A diagram of the temperature (left) and $CO_2$  (right) control systems.}
\label{diagram}
\end{figure*}

The temperature control system (Fig. \ref{diagram}) begins with an analog temperature sensor (TMP37FT9Z, Analog Devices) which is monitored via an analog input on the Arduino. Input voltages from the temperature sensor are converted to degrees Celsius. Optimally, mammalian cells like to be held at $37^oC$, so when the temperature measurement drops below a threshold of $35^oC$, a mini, corded desktop heater (WY-H1, Ucan) plugged into a 110V outlet is switched on via a 5V relay module (SRD-05VDC-SL-C, Tolako) controlled by one of the digital outputs of the Arduino.\cite{outlet} The digital output to the relay module is held high while the temperature in the incubator increases and doesn't switch to low until the temperature exceeds $37^oC$. At that point, power to the heating unit is turned off. The internal temperature of the imaging incubator was confirmed via a hand held infrared thermometer (Lasergrip 1080, Etekcity). The desired temperature of $37^oC$ was used as the upper threshold to turn off the heating unit because the ambient heat built up in the unit will still dissipate within the enclosure, causing the internal temperature to continue to rise. Setting a higher threshold than $37^oC$ allowed temperatures to climb beyond the desired temperature range. Keeping the $35^oC$ and $37^oC$ lower and upper thresholds respectively gave a stable temperature range, keeping the temperature system from cycling excessively. Excessive cycling of the heating unit was also avoided by using the average of one hundred temperature measurements in quick succession, thereby mitigating fluctuations in the temperature measurements due to noise within the system.

\begin{figure}[htb]
\centering\includegraphics[width=8cm]{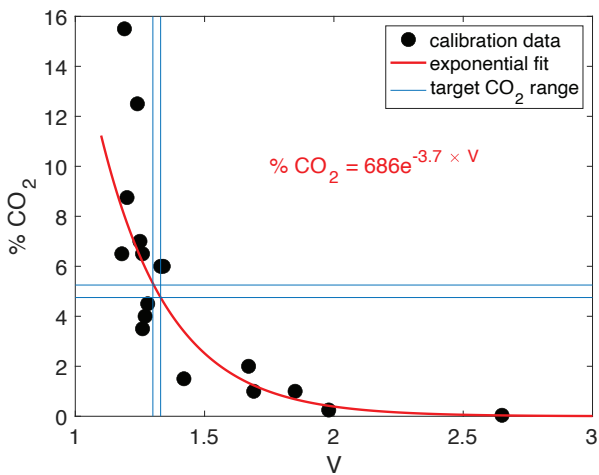}
\caption{Calibration of the analog $CO_2$ sensor.}
\label{calibration}
\end{figure}

Similar to the temperature control system, the $CO_2$ control system (Fig. \ref{diagram}) has an analog $CO_2$ gas sensor (SEN0159, DFROBOT) connected to an analog input on the Arduino.\cite{CO2} The $CO_2$ sensor needs to be calibrated to set the optimal range for $CO_2$ concentrations---typically $5\% CO_2$ for mammalian cells. The analog device was calibrated using a fyrite gas analyzer kit (Model 10-5000, Bacharach), borrowed from the university's Biology Department. The analog output of the device decreases along an exponential decay function as $CO_2$ concentrations increase (Fig. \ref{calibration}). When $CO_2$ levels drop below the lower threshold of $4.8\% CO_2$, the Arduino triggers the activation of a 12V two-way normally closed electric solenoid air valve (2V025-08, AOMAG), which allows pressurized $CO_2$ to supply the imaging incubator. When the concentration of $CO_2$ climbs above the upper threshold of $5.2\% CO_2$, the solenoid is deactivated, thereby closing its valve and ceasing the supply of $CO_2$ to the incubator. The solenoid is the only major electrical component of the imaging incubator that is not housed within the enclosure. As with the temperature control system, $CO_2$ concentrations were determined by averaging one hundred measurements in quick succession; the lower and upper thresholds used here gave a stable range of $CO_2$ concentrations, keeping the $CO_2$ control system from cycling excessively.

A 75 lb $CO_2$ tank in the lab that is used for a standard tissue culture incubator was equipped with a two-way gas manifold in order to supply the imaging incubator simultaneously. The $CO_2$ port lies above the sample stage, which directed pressurized gas at the sample during imaging. In order to prevent the pressurized gas from perturbing the sample or increasing the loss of media, an additional length of tubing was added within the imaging incubator in order to redirect the delivery of $CO_2$ away from the sample stage. In addition, the placement of the tubing's output nozzle was placed near the heating unit's fan in order to promote mixing of $CO_2$ throughout the imaging incubator.

While the Arduino can be powered via the USB 2.0 cable that links it to the computer, the many devices controlled via the Arduino tax its maximum output. In particular, the solenoid that serves as the automated valve for $CO_2$ injection requires greater current and voltage outputs than can be provided by the Arduino alone. An external power cable connects the Arduino directly to an external electrical outlet, which supplies additional power for the devices via the $V_{cc}$ pin. Even with the external power supply to the Arduino, a Darlington transistor (TIP120, onsemi) is needed in order to amplify the input to the solenoid (Fig. \ref{diagram}).\cite{solenoid}

\subsection{Fluid System}
In order to keep cells viable during imaging, they must also be submerged under adequate media as both a nutrient source and a means of preventing cellular dehydration. An imaging chamber (QR-40HP, Warner instruments) holds the coverslip to which cells are adhered, along with a few mL of media. Note that the cost of the imaging incubator noted in the introduction does not include that of the imaging chamber, which was already in the lab's possession. Similar imaging chambers could be made in-house via a 3D printer in order to save costs.

A water bath is generally used at the base of standard tissue culture incubators in order to keep humidity levels high an prevent media from evaporating. The imaging incubator does not have this luxury, as increased humidity within the imaging incubator would increase the likelihood of condensation building upon optical elements, thereby degrading image quality. Furthermore, increased humidity within the imaging incubator could also affect the functionality of electronic components housed within the imaging incubator. Evaporation of media within the imaging incubator is unavoidable given the internal heating unit and temperature requirements to maintain cell viability. Evaporation of media in the imaging incubator has two negative side effects, the first being the possible build up of condensation on optical elements as already mentioned and the second being the depletion of cells' media. Media that has evaporated away from the cell imaging chamber is then replenished via a gravity fed 15 drops/mL I.V. drip system, with the drip rate controlled using a variable latch.

\section{Software Design}
\label{software}
System automation was an essential requirement of the system's software design; our goal of keeping cells viable for 24 hours of imaging meant that the system would be left unattended for extended periods of time. As such, we decided to control the Arduino---and therefore the associated hardware devices---using LabVIEW and the LINX add-on package for LabVIEW.\cite{LINX} This way, the imaging incubator and imaging device programming and automation could be integrated for ease of use. As an alternative, the Arduino IDE could also have been used to program the imaging incubator separately from the imaging system program. But again, we opted to run a single program instead of having two separate programs competing for computational resources.

To initialize the Arduino for LabVIEW, the MakerHUB LINX Firmware Wizard was used to install the appropriate driver. Then in LabVIEW, communication with the Arduino is made simple via the LINX add-on. To begin, communication is set up with the Arduino via the Open.vi provided by the LINX add-on---the .vi file extension stands for Virtual Instrument in LabVIEW.

The program first checks the temperature within the imaging incubator using the TMP3x.vi provided by the LINX add-on. The TMP3x.vi is written to operate three different temperature sensors (the TMP35, TMP36 and TMP37), so the appropriate sensor needs to be selected in the drop down provided (in this case the TMP37) in order to call to the appropriate calibration. The TMP3x.vi is a modification of the Analog Read.vi provided by the LINX add-on in order to read input voltages via the Arduino. With any of the analog or digital read/write VI's in the LINX add-on, the appropriate Arduino pin number corresponding to the wired component(s) must be provided. When temperatures outside of the range of the lower or upper temperature thresholds are read by the Arduino, the program then triggers the appropriate on or off state of a digital Arduino pin via the Digital Write.vi, thereby controlling the 5V relay module and by it, the external power outlet supplying the heating unit (Fig. \ref{diagram}).

The $CO_2$ control system first calls to the Analog Read.vi in order to read the voltage input from the $CO_2$ sensor via the Arduino. When the soleniod needs to be opened or closed in order to control $CO_2$ injection, the Digital Write.vi is implemented in order to turn the base of the TIP120 Darlington transistor on or off via a high or low digital output, respectively (Fig. \ref{diagram}). When the digital output connected to the base is active, the current gain of the Darlington transistor is sufficient to power the solenoid.

\begin{figure}[htb]
\centering\includegraphics[width=8.5cm]{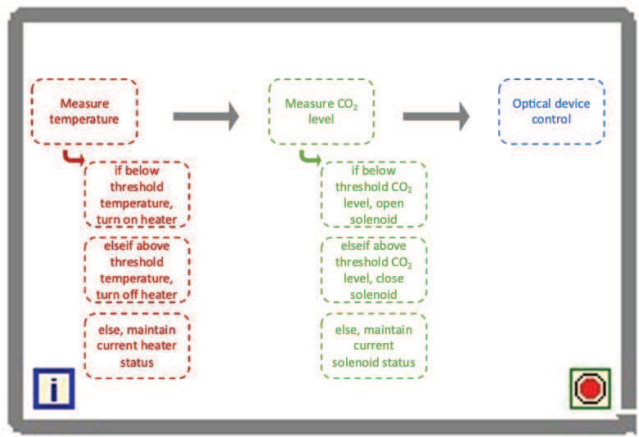}
\caption{A programmatic flow diagram for the systems control of the imaging incubator. Each system is called to sequentially within a while loop such that the systems are continually being called to and operated on throughout imaging.}
\label{flow}
\end{figure}

After controlling the environmental systems of the imaging incubator, the program then goes on to control the imaging system itself (not detailed here). The control sequencing (Fig. \ref{flow}) is contained within a while loop with a 1 second delay between iterations so as to prevent the control systems from cycling excessively. The while loop maintains system operations until the system is either halted manually via a stop button on the program front panel, or until the experiment is completed. At that point, all digital outputs from the Arduino are set to low via the Digital Write.vi so as to turn off the respective devices and communications with the Arduino are terminated via the Close.vi, again provided by the LINX add-on.

The system was run using a Dell Precision T3500 workstation (64-bit, $\times 8$ Intel Xeon CPU W3550 @ 3.07GHz, 6.00 GB RAM, Windows 10 Pro).

\section{Results}
\label{results}
\begin{figure}[htb]
\centering\includegraphics[width=8.5cm]{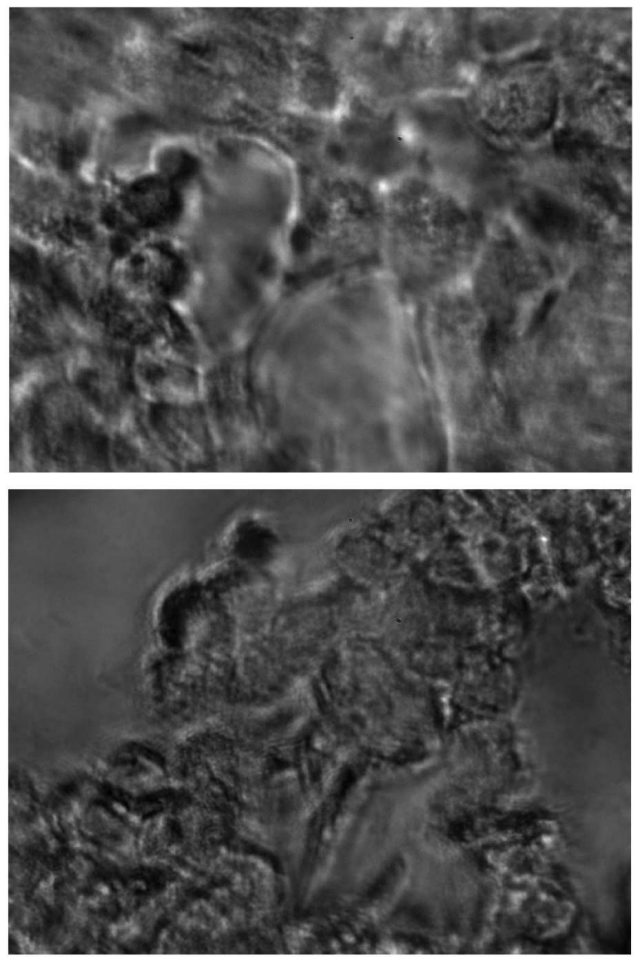}
\caption{Living T-47D breast cancer cells imaged over 24 hours in the Arduino-based imaging incubator. (Top) Initial image acquired at the start of imaging. (Bottom) Final image acquired after 24 hours of imaging. (Note, the initial and final fields of view displayed are different because the imaging chamber was bumped during imaging when wicking off excess media while adjusting the drip system flow rate.)}
\label{cells}
\end{figure}

As a proof of the system's functionality, live imaging of human breast cancer cells (T-47D, ATCC) was conducted over 24 hours while in the Arduino-based imaging incubator (Fig. \ref{cells}) . A $\#1$ glass coverslip with cells adhered was held in the imaging chamber with 3 mL of Dulbecco's modified Eagle's medium (DMEM, HyClone) with 10\% fetal bovine serum (Gibco), 100 IU/mL penicillin and 100$\mu$g/mL streptomycin added (MP Biomedicals). Media was further supplied throughout imaging via the gravity-fed drip system. A period of approximately 50 seconds between drips seems to be a decent rate for replenishing the imaging chamber with media---neither too great so as to overfill the well, nor so slow so as to allow the media to evaporate off completely. No condensation was observed within the imaging incubator during the 24 hour imaging session.

Cells were imaged via a custom white-light imaging microscope in transmission using a 40X microscope objective (PLN 40X, Olympus). Images were acquired using a monochromatic InGaAs camera (Alvium 1800 U-319m, Allied Vision) at 10 minute intervals over the 24 hour imaging period.

\section{Conclusion}
\label{conclusion}
An Arduino-based, low-cost imaging incubator was developed in order to maintain viable cells for extended periods of time while imaging via a custom imaging device. Use of the imaging incubator can allow for real-time imaging of cells while undergoing various stimuli or perturbations. Moreover, real-time imaging of viable cells during experimental protocols allows researchers to investigate the time-dependent response of cells to stimuli without the need for repeated experiments at multiple, discrete time points and without the need to repeatedly transfer cells back and forth between incubators and the imaging device. The repeated transfer of cells between incubators and imaging devices opens the cells to outside factors that cannot be controlled for, such as environmental factors or threat of infection.

The Ardunio-based imaging incubator presented here gives a low-cost alternative to costly stage-top incubators that can be purchased for imaging applications via scientific retailers. All of the materials and equipment used to fabricate the Arduino-based imaging incubator were purchased for under \$350 and could be purchased from local hardware, online and maker retailers, providing researchers savings of thousands to tens of thousands of dollars in comparison to devices with similar features available from scientific retailers. The versatility of the imaging incubator presented here is also an advantage of the system, as it can be adapted to suit standard microscopes as well as any and all custom imaging devices. The imaging incubator presented would be applicable for all conceivable imaging applications, including scatter imaging, holography, differential interference contrast microscopy and even fluorescence microscopy. The later assumes cells have been transfected with fluorescent label(s) for live-cell fluorescence imaging. Images of live cells over the course of 24 hours while in the Arduino-based imaging incubator demonstrate the system's functionality.

\section*{Appendix A: Component List}
\label{appendix}
Components used to construct the imaging incubator are listed below. Specific part numbers and manufacturers are listed where appropriate. Part numbers are not specified for generic parts that can be purchased from a local hardware store or retailer, due to variability in local suppliers and in the required enclosure size. Single item quantities are assumed unless otherwise specified.\\

Enclosure:
\begin{itemize}
	\setlength\itemsep{0em}
	\item Clear acrylic plexiglass sheets ($\times5$, 1/8'' thick, size depends upon system dimensions)
	\item Right-angle brackets ($\times8$) with associated screws and nuts
	\item Light-weight cabinet hinges ($\times2$) with associated screws and nuts
	\item 1/2'' water tight conduit connectors ($\times2$)
	\item 1/4'' hose barbs ($\times2$, male threaded fittings)
	\item Female-female threaded adapter to join the two hose barbs
	\item O-rings, sized to fit the conduit connectors and hose barbs
	\item Silicone rubber sheet (1/16'' thick) cut into 1'' wide strips of appropriate lengths
	\item Clear caulking
	\item Cabinet knob
	\item Magnetic cabinet door catch
	\item UV lamp (CTUV-6WTIMERUVC, Aopu Lighting Co)
\end{itemize}

Temperature Control System:
\begin{itemize}
	\setlength\itemsep{0em}
	\item Arduino Uno with protoboard and external power supply (the same Arduino will also used for the $CO_2$ control system)
	\item Analog temperature sensor (TMP37FT9Z, Analog Devices)
	\item Mini corded desktop heater (WY-H1, Ucan)
	\item 5V relay module (SRD-05VDC-SL-C, Tolako)
	\item 15A grounding duplex outlet
	\item Duplex outlet wall plate
	\item 1-gang non-metallic weatherproof box
	\item 3/8'' non-metallic twin-screw cable clamp connector
	\item 14 AWG power cord with grounding plug
	\item Miscellaneous jumper wires
\end{itemize}

$CO_2$ Control System:
\begin{itemize}
	\setlength\itemsep{0em}
	\item Analog $CO_2$ gas sensor (SEN0159, DFROBOT)
	\item 12V two-way normally closed electric solenoid air valve (2V025-08, AOMAG)
	\item 1/4'' hose barbs ($\times2$) for input/output from solenoid
	\item Darlington transistor (TIP120, onsemi)
	\item 1N4001 Diode
	\item $1k\Omega$ resistor
	\item $CO_2$, can be supplied by tank (may need a two-way gas manifold) OR from the building's internal supply
	\item 1/4'' tubing, length to be determined by lab space
	\item Miscellaneous jumper wires
\end{itemize}

Fluid System:
\begin{itemize}
	\setlength\itemsep{0em}
	\item Imaging chamber (not included in total pricing of imaging incubator components, various options are available on the market, but we used the QR-40HP, Warner instruments, MA)
	\item 15 drops/mL I.V. drip system
	\item Lure lock
	\item 60mL syringe, used as media reservoir, attached to I.V. drip tubing via lure lock
	\item Hypodermic needle, used to better direct media drips to the imaging chamber
	\item Buret clamp, to hold syringe elevated above imaging incubator
	\item Miscellaneous clamps and hardware for mounting hypodermic needle above imaging chamber but outside of FOV
\end{itemize}

\section*{Appendix B: LabVIEW Block Diagrams}
\begin{figure*}[htb]
\centering\includegraphics[width=17cm]{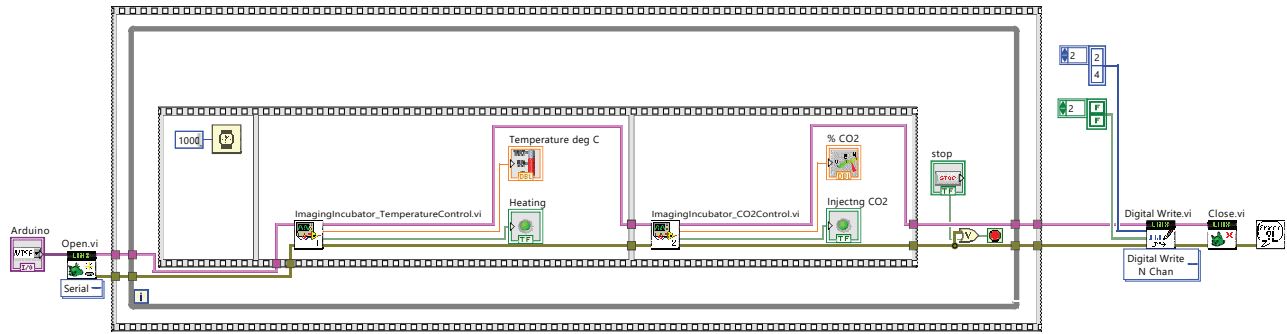}
\caption{The main block diagram for control of the Arduino-based imaging incubator via LabVIEW. Specific VI's are labeled in the diagram, as detailed in Sec. \ref{software}. This block diagram is available online as Supplement 1.}
\label{BlockDiagram}
\end{figure*}

\begin{figure*}[htb]
\centering\includegraphics[width=17cm]{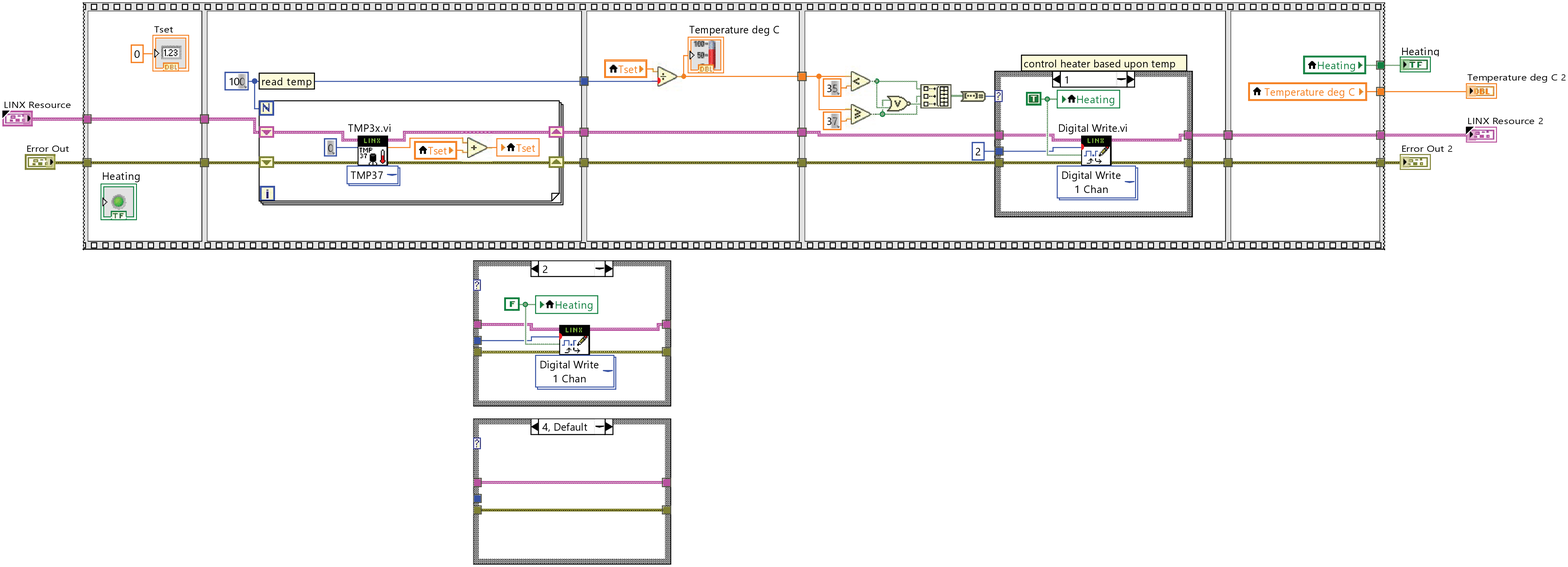}
\caption{The block diagram for the ImagingIncubator\textunderscore TemperatureControl.vi sub-VI within the main program. Hidden cases within the case structure are displayed separately below the main block diagram. Specific VI's are labeled in the diagram, as detailed in Sec. \ref{software}. This block diagram is available online as Supplement 2.}
\label{BlockDiagramTemp}
\end{figure*}

\begin{figure*}[htb]
\centering\includegraphics[width=17cm]{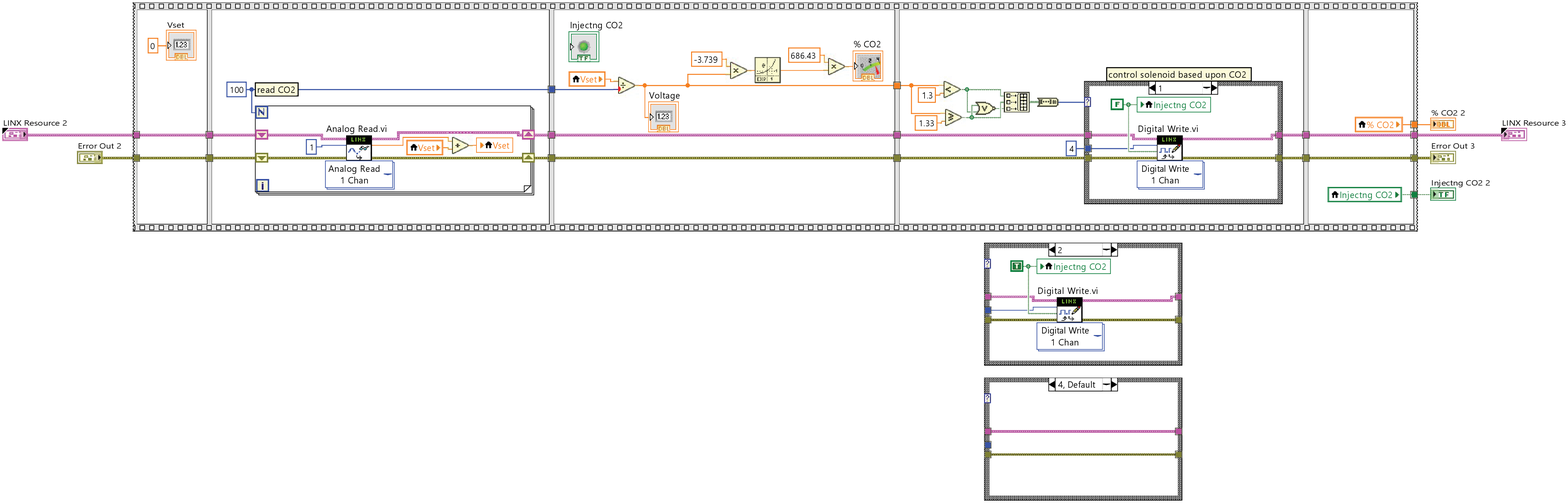}
\caption{The block diagram for the ImagingIncubator\textunderscore CO2Control.vi sub-VI within the main program. Hidden cases within the case structure are displayed separately below the main block diagram. Specific VI's are labeled in the diagram, as detailed in Sec. \ref{software}. This block diagram is available online as Supplement 3.}
\label{BlockDiagramCO2}
\end{figure*}

LabVIEW block diagrams are included for control of the Arduino-based imaging incubator. Imaging controls are not included as they are beyond the scope of this paper. To aid in printability, the temperature and $CO_2$ control systems are written as sub-VI's within the main VI (Fig. \ref{BlockDiagram}) and are given in Figs. \ref{BlockDiagramTemp} and \ref{BlockDiagramCO2}, respectively. In order to make the figures accessible, they are also included as separate PDF files in the article's online supplemental materials.

\begin{backmatter}
\bmsection{Funding} This project was supported by an Institutional Development Award (IDeA) from the National Institute of General Medical Sciences of the National Institutes of Health under grant number P20 GM103418.

\bmsection{Disclosures} The authors declare no conflicts of interest.

\bmsection{Data availability} No data were generated or analyzed in the presented research.

\bmsection{Supplemental document} See Supplements 1-3 for supporting content. 

\end{backmatter}

\bibliography{ImagingIncubator}






\end{document}